\documentclass[11pt]{article}
\usepackage{epsfig}
\newcommand{\be}{\begin{equation}}
\newcommand{\ee}{\end{equation}}
\newcommand{\bra}{\langle}
\newcommand{\ket}{\rangle}
\newcommand{\bea}{\begin{eqnarray}}
\newcommand{\eea}{\end{eqnarray}}
\newcommand{\DT}{\Delta t}

\begin{document}

\title{Testing and tuning symplectic integrators for Hybrid Monte Carlo algorithm in lattice QCD}

\author{Tetsuya Takaishi${}^{a}$ and Philippe de Forcrand${}^{b,c}$ \\
\small \it $^a$Hiroshima University of Economics, Hiroshima 731-0124, Japan \\
\small \it $^b$CERN, Physics Department, TH Unit, CH-1211 Gen\`eve 23, Switzerland \\
\small \it $^c$Institute for Theoretical Physics, ETH Z\"urich, CH-8093 Z\"urich, Switzerland  \\
}

\maketitle

\abstract{ 
We examine a new 2nd order integrator recently found by Omelyan et al.
The integration error of the new integrator  measured in the root mean square of the energy 
difference, $\bra\Delta H^2\ket^{1/2}$, is about 10 times smaller 
than that of the standard 2nd order leapfrog (2LF) integrator.
As a result, the step size of the new integrator can be made about three times larger.
Taking into account a factor 2 increase in cost, the new integrator
is about 50\% more efficient
than the 2LF integrator.  
Integrating over positions first, then momenta, is slightly more advantageous than the reverse.
Further parameter tuning is possible. We find that the optimal parameter for the new integrator 
is slightly different from the value obtained by Omelyan et al., and depends on the
simulation parameters.  
This integrator
could also be
advantageous for the Trotter-Suzuki decomposition in Quantum Monte Carlo.
}

\section{Introduction}

The Hybrid Monte Carlo (HMC) algorithm\cite{HMC} is now the established
standard for the generation of dynamical fermion configurations in lattice Quantum Chromo Dynamics (QCD).
The HMC algorithm consists of molecular dynamics (MD) trajectories, each followed by a
Metropolis test.
During the MD trajectory, one integrates Hamilton's equations of motion,
using an integrator with a  discrete stepsize $\Delta t$ which must satisfy two conditions
in order to maintain detailed balance:
$(i)$ simplecticity (the phase space volume $dp dq$ must be conserved) 
and $(ii)$ time reversibility. 
The simplest and most widely used integrator wih these properties
is the 2nd order leap frog (2LF) integrator,
which causes ${\cal O}(\Delta t^2)$ errors in the total energy or Hamiltonian. 
These errors are eliminated at the Metropolis accept/reject step,
which makes the algorithm exact. 

The acceptance at the Metropolis step depends on the magnitude of 
the error in the total energy.
In order to reduce the error and thus increase the acceptance 
one could use higher order integrators. 
Early attempts, however,
did not appear to be practical \cite{Campostrini:1989ac,Creutz}.
This is because the efficiency of higher order integrators depends largely on 
the system size and these early attempts were made on rather small lattices. 
As the lattice size increases above a certain value $V_c$,
the higher order integrators should perform better than the low order integrator.
This minimum lattice size $V_c$ depends on the Hamiltonian which we consider and 
on the choice of integrator.
For lattice QCD, it turned out that $V_c$ becomes very large at small quark masses,
so that on currently accessible computers the 2LF integrator is the best choice\cite{Takaishi1}
for simulations at zero temperature.
At finite temperature, higher order integrators could perform better on moderate-size lattices\cite{Takaishi2}: this is because chiral symmetry gets restored,
so that small Dirac eigenvalues disappear, which allows for stable MD integration
using larger stepsizes.

So far, only the 2LF integrator has been considered in the HMC algorithm of lattice QCD as a second order integrator,
because of its simplicity and effectiveness.
Recently however, Omelyan {\it et al.} \cite{Omelyan} found a new 2nd order integrator which 
is expected to be better than the 2LF integrator although it
has twice the computational cost. 
Here we examine this new 2nd order integrator for the HMC algorithm in lattice QCD 
and measure its efficiency.
We also examine the new 4th order integrators recommended in \cite{Omelyan}.
Finally, we try to further tune the new 2nd order integrator.

\section{Symplectic Integrator}
\subsection{Recursive construction scheme}
Symplectic integrators are most conveniently described by 
the Lie algebra formalism\cite{Forest:1989ez,Yoshida,Sexton:1992nu}.
Let $H$ be a classical Hamiltonian,
\be
H=\frac12 p^2 +S(q),
\label{Hamiltonian}
\ee
where $q=(q_1,q_2,\dots)$ and $p=(p_1,p_2,\dots)$ are coordinate variables and conjugate momenta respectively.
Hamilton's equations are expressed as
\be
\dot{f}=\{f,H\},
\label{eq:Hamilton}
\ee
where $f= q$ or $p$ and $\{,\}$ stands for the Poisson bracket,
\be
\{f,H\}=\sum_i \left(\frac{\partial f}{\partial q_i}\frac{\partial H}{\partial p_i}
-\frac{\partial f}{\partial p_i}\frac{\partial H}{\partial q_i}\right).
\ee
If we define the linear operator $L(H)$ as
\be 
L(H)f \equiv \{f,H\},
\ee
then we can write the formal solution of Hamilton's equations,
\be
f(t+\Delta t)=\exp(\Delta t L(H))f(t).
\ee
In general the operator $\exp(\Delta t L(H))$ cannot be expressed exactly in a
simple form.
Therefore we approximate $\exp(\Delta t L(H))$ 
with an operator correct up to a certain order in $\Delta t$.
Let us write $L(H)$ as
\bea
L(H)& = & L(\frac12 p^2) +L(S(q)), \\
    & = & T+ V,
\eea 
where $T \equiv L(\frac12 p^2)$ and $V\equiv L(S(q))$.
The 2LF integrator is given by 
decomposing $e^{\Delta t (T+V)}$ as
\be 
\exp(\Delta t~(T+V)) = \exp(\frac12 \Delta t~T)\exp(\Delta t~V)\exp(\frac12 \Delta t~T) +{\cal O}(\Delta t^3).
\ee
We call $G_{2}(\Delta t)$ the 2LF integrator:
\be 
G_{2}(\Delta t) \equiv \exp(\frac12 \Delta t~T)\exp(\Delta t~V)\exp(\frac12 \Delta t~T).
\label{G_2}
\ee
The integrator $G_{2}(\Delta t)$ amounts to mapping $q$ and $p$ to new variables as

\bea 
\left( \begin{array}{c}
q(t+\DT) \\
p(t+\DT)
\end{array} \right)
& = &
\left( \begin{array}{cc}
1 & \frac12 \DT \\
0 & 1  
\end{array} \right)
\bullet
\left( \begin{array}{cc}
1 & 0  \\
- \frac{\partial S(q)}{\partial q}\frac{\DT}{q} & 1  
\end{array} \right)
\bullet
\left( \begin{array}{cc}
1 & \frac12 \DT \\
0 & 1  
\end{array} \right)
\bullet
\left( \begin{array}{c}
q(t) \\
p(t)
\end{array} \right) \\ 
& \equiv &
G_2(\DT)\bullet 
\left( \begin{array}{c}
q(t) \\
p(t)
\end{array} \right).
\eea
 
This map is symplectic. This is easy to see, since the three matrices
representing the elementary substeps are triangular with determinant 1.
It is also exactly time-reversible: $G_2(\DT)G_2(-\DT)=1$. 

An equivalent algorithm is obtained by interchanging $T$ and $V$ in eq.(\ref{G_2}).

Higher order integrators can also be found by decomposing  $e^{\Delta t (T+V)}$ to the desired order.
Although the decomposition to a higher order is a non-trivial problem with no
unique solution, there is a simple recursive construction scheme which 
generates higher order integrators from lower order ones\cite{Creutz,Suzuki,Yoshida}.
In this scheme, the $(2k+2)$-th order integrator is given by
\be 
G_{2k+2}(\Delta t)=G_{2k}(b_1 \Delta t)G_{2k}(b_2 \Delta t)G_{2k}(b_1 \Delta t),
\label{eq:rec}
\ee
where
\be
b_1 = \frac1{2-2^{1/(2k+1)}},
\ee
\be
b_2 =1-2b_1 = - \frac{2^{1/(2k+1)}}{2-2^{1/(2k+1)}}.
\ee
Let us call the integrators of eq.(\ref{eq:rec}) recursive construction (RC) integrators. 
These integrators are symplectic and constructed in a symmetric way, thus time reversible, 
i.e. $G_{2k+2}(\Delta t)G_{2k+2}(-\Delta t)=1$.
Note that $b_2$ is negative.
The appearance of negative coefficients in higher order integrators is inevitable: beyond the 2nd order decomposition 
there is no decomposition scheme having positive coefficients only\cite{Suzuki2}.
However if we include the commutator $[V,[T,V]]$ in the decomposition
we may circumvent this situation\cite{Suzuki3,Chin},
and integrators with all positive coefficients can be constructed. 
Even so, the inclusion of this commutator requires the calculation of the gradient of the force,
which increases the computational cost.  
If the force-gradient  calculations are computationally simple for the system considered,
then it would be worth considering such integrators.
For lattice QCD, it is unclear that such force-gradient integrators have advantages over the non-force-gradient ones.
We do not consider such force-gradient integrators here.

\subsection{Minimum norm construction scheme}

Although the recursive construction scheme makes it easy to construct higher order integrators to any order,
their performance may not be optimal, since the number of force calculations grows rapidly with the order
of the integrator. 
More generally, one can decompose $e^{\Delta t (T+V)}$ as
\be 
\exp(\DT~(T+V)) =
\Pi_i^k \exp(c_i \Delta t~T)\exp(d_i \Delta t~V) +{\cal O}(\Delta t^{n+1}),
\ee
where $\sum_i^k c_i =\sum_i^k d_i =1$.
Moreover, in order to form a time-reversible integrator 
certain relations must be hold.
For instance if we take $k=3$, 
the following equations must be satisfied:
$(i)$$c_1=c_4, c_2=c_3,d_1=d_3,d_4=0$ or $(ii)$$c_1=0, c_2=c_4,d_1=d_4,d_2=d_3$.
For time-reversible integrators, the error terms with odd $n$ always 
vanish\cite{Creutz,Yoshida,Suzuki2}. 
Thus time-reversible integrators have a leading error term ${\cal O}(\Delta t^{n+1})$ 
with $n$ even.

The error term ${\cal O}(\Delta t^{n+1})$ consists of commutators of $T$ and $V$.
For instance, the leading error terms of the 2nd and 4th order integrators are 
respectively \cite{Omelyan}
\be
{\cal O}(\Delta t^{3})= \alpha [T,[V,T]] + \beta [V,[V,T]],
\label{eq:alpha}
\ee
and
\be
{\cal O}(\Delta t^{5})= \gamma_1 [T,[T,[T,[T,V]]]]+\gamma_2 [T,[T,[V,[T,V]]]]+\gamma_3[V,[T,[T,[T,V]]]]
+\gamma_4 [V,[V,[T,[T,V]]]],
\label{eq:gamma}
\ee
where $\alpha$, $\beta$ and $\gamma_i$ depend on $c_i$ and $d_i$.

One strategy to find optimal integrators in the absence of further information
about the operators $T$ and $V$ is to minimize the norm of the error coefficients.
For the case of eq.(\ref{eq:alpha}) and (\ref{eq:gamma}), this strategy implies minimizing the following error functions:   
\be 
Err_3 \equiv \sqrt{\alpha^2 + \beta^2},
\label{eq:Err3}
\ee
and
\be
Err_5 \equiv \sqrt{\gamma_1^2 + \gamma_2^2 +\gamma_3^2 +\gamma_4^2 }.
\label{eq:Err5}
\ee
Omelyan {\it et al.} \cite{Omelyan} found a class of integrators by following
this strategy\footnote{Note that in some cases, the set
of polynomial equations defining the optimal decomposition can be solved
analytically, even beyond the second-order case \cite{ADK}.}.
Among the new integrators which they identified, 
they found several ``outstanding'' integrators having 
especially small norms of the error coefficients. 
In this analysis, we consider the new 2nd and 4th order integrators which they recommend as  outstanding integrators,
and which are described as follows.

\subsubsection{2nd order minimum norm (2MN) integrator}

Omelyan {\it et al.} \cite{Omelyan,Omelyan2} obtained the following new 2nd order integrator.
\be
I_{2MN}(\DT)=e^{\lambda \DT T}e^{\frac{\DT}{2}V}e^{(1-2\lambda) \DT T} e^{\frac{\DT}{2}V}e^{\lambda \DT T},
\label{eq:2MN}
\ee
where $\lambda$ takes value $\lambda_c$:
\be
\lambda_c = \frac12 -\frac{(2\sqrt{326}+36)^{1/3}}{12}+\frac1{(6\sqrt{326}+36)^{1/3}} \approx  0.1931833275037836  .
\label{eq:optlambda}
\ee
This value of $\lambda$ minimizes 
$\displaystyle \sqrt{\alpha(\lambda)^2 + \beta(\lambda)^2}$, where
\be
\alpha(\lambda) = \frac{1-6\lambda +6\lambda^2}{12},
\label{eq:alpha_lambda}
\ee
\be
\beta(\lambda) = \frac{1-6\lambda}{24}
\label{eq:beta_lambda}
\ee
as can be derived from the expansion of (\ref{eq:2MN}).

This integrator requires two force calculations per step.
Thus, compared to the 2LF integrator, it has twice the computational cost.
The norm of the error coefficients $Err_3$, however, is a factor of 10 smaller
 ( $Err_3^{2LF}/Err_3^{2MN}\approx 10.9 $\cite{Omelyan2} ).
As we will see later, the error of a 2nd order integrator 
at the end of a Hybrid Monte Carlo trajectory
is expected to be proportional to $\DT^2$.
Therefore, even after taking into account the increased computational cost,
we expect that  the 2MN integrator will perform better than the 2LF integrator,
by a factor $\approx \sqrt{10.9}/2$.
We will numerically confirm this in the next section,
and later we will further try to tune the integrator by modifying the error function.

\subsubsection{ 4th order minimum norm (4MN) integrator }

At the beginning of the MD integration one can start the integration with either $q$ or $p$.
Usually we do not consider this freedom seriously since  
for the 2nd order integrator the choice of the starting variable 
does not make a significant difference in performance\footnote{Actually there is a small difference, which we identify later.}.
In general, however, the performance could be different 
depending on the choice of the starting variable.
In fact, the optimal integrator itself could also be different 
depending on the starting variable. 

This is precisely what Omelyan {\it et al.} found for higher order MN integrators.
Let us call {\it velocity version}\footnote{One could say {\it momentum version}. 
Here we follow the convention in the literature.} the integrator starting by integrating $p$
and {\it position version} the integrator starting by integrating $q$.
For the optimal 4th order MN integrators with 5 force calculations 
they found that the velocity version has smaller errors than  the position version.
Actually we have tested both integrators and also found that typically the error of the velocity version 
is a few times smaller than that of the position version.
In the following numerical tests we use only the velocity version of 
the 4th order MN integrators with 5 force calculations (4MN5FV)  
which is written as\cite{Omelyan}
\be
I_{4MN5FV}(\DT)=e^{\theta \DT V}e^{\rho \DT T}e^{\lambda \DT V}e^{\mu \DT T}e^{(1-2(\lambda+\theta)) \frac{\DT}2 V}e^{(1-2(\mu+\rho)) \DT T}
e^{(1-2(\lambda+\theta)) \frac{\DT}2 V}e^{\mu \DT T}e^{\lambda \DT V}e^{\rho \DT T}e^{\theta \DT V},
\ee 
where
\bea
\theta & = & 0.08398315262876693 \\ \nonumber
\rho  & = & 0.2539785108410595 \\ \nonumber
\lambda  & = & 0.6822365335719091  \\ \nonumber
\mu  & = & -0.03230286765269967.
\eea

Furthermore we have also tested the position version of 
the 4th order MN integrators with 4 force calculations (4MN4FP) 
given by\cite{Omelyan}
\be
I_{4MN4FP}(\DT)=e^{\rho \DT T}e^{\lambda \DT V}e^{\theta \DT T}e^{(1-2\lambda)\frac{\DT}2 V}
e^{(1-2(\theta+\rho)\DT T}e^{(1-2\lambda)\frac{\DT}2 V}e^{\theta \DT T}e^{\lambda \DT V}e^{\rho \DT T},
\ee
where
\bea
\rho & = &  0.1786178958448091 \\ \nonumber
\theta  & = & -0.06626458266981843\\ \nonumber
\lambda  & = & 0.7123418310626056  .
\eea
The velocity version (4MN4FV) is expected to have a similar error to the 
position version 4MN4FP above\cite{Omelyan}.
Thus we used the 4MN4FP integrator which has one less force evaluation per MD trajectory.

\section{Numerical tests of the new integrators}
\subsection{Lattice QCD action}
We use the plaquette Wilson gauge and standard Wilson fermion actions
with two degenerate fermion flavors\cite{Wilson}. 
The partition function is given by
\be 
Z=\int {\cal D}U\det [M(U)M(U)^\dagger]\exp(-S_g(U)),
\label{eq:partition}
\ee
with $S_g(U)$ the gauge action given by
\be
S_g(U)=\frac{\beta}{3}\sum_{U_p} Tr [1-U_p(U)],
\ee
where $U_p(U)$ stands for the plaquette, $\beta$ is the gauge coupling
and $U$ is an SU(3) link variable.
$M(U)$ is the Wilson Dirac operator
defined by
\be
M_{ij}(U)=\delta_{i,j}+\kappa \sum_{\mu}[(\gamma_\mu-1) U_{i,\mu}\delta_{i,j+\mu}
-(\gamma_\mu +1)U^{\dagger}_{i-\mu,\mu}\delta_{i,j+\mu}],
\label{eq:Wilson}
\ee
where $\kappa$ is the hopping parameter and $\gamma_\mu$ are the $\gamma$ matrices.
The inverse of the hopping parameter $\kappa$ is related to the quark mass $m$  
and the value $\kappa$ where $m$ is zero is denoted by $\kappa_c$.
As $m$ decreases  the eigenvalues of the matrix $M$ 
become small and at the zero quark mass limit   
the matrix $M$ becomes singular.

Using pseudofermion fields $\phi$ the partition function is re-expressed as
\be
Z = \int {\cal D}U {\cal D}\phi^\dagger {\cal D}\phi \exp[-\phi^\dagger (M(U)M(U)^\dagger)^{-1}\phi-S_g(U)].
\ee
Furthermore introducing momenta $p$, we obtain
\be
Z = \int {\cal D}U {\cal D}\phi^\dagger {\cal D}\phi {\cal D}p~\exp[-H(U,\phi,\phi^\dagger,p)],
\ee
\label{eq:MDHamiltonian}
where 
\be
H(U,\phi,\phi^\dagger,p) = \frac{1}{2}\sum p^2 + \phi^\dagger (M(U)M(U)^\dagger)^{-1}\phi + S_g(U)
\ee
is the Hamiltonian we consider in our numerical tests.

\subsection{Hybrid Monte Carlo algorithm}
The HMC algorithm combines MD and Metropolis accept/reject steps\cite{HMC}
to form a Markov chain.
Starting from an ``old'' configuration $\{U\}$, a ``candidate'' configuration
$\{U'\}$ is obtained by 
$(i)$ drawing momenta $\{p\}$ and pseudofermion fields $\{\phi\}$ from
Gaussian distributions $\exp(-\frac{1}{2}p^2)$ and 
$\exp(-\phi^\dagger (M(U)M(U)^\dagger)^{-1}\phi)$ respectively;
$(ii)$ integrating Hamilton's equations of motion eq.(\ref{eq:Hamilton}) 
with a discrete stepsize integrator.
In order to maintain detailed balance  
the integrator in the MD step must satisfy two conditions: simplecticity  and time reversibility.
Let $T_{MD}(\Delta t)$ be an elementary MD step with a discrete step size $\Delta t$.
$T_{MD}(\Delta t)$ evolves $(p,q)$ to $(p^\prime,q^\prime)$:
\be 
T_{MD}(\Delta t):(p,q)\rightarrow (p^\prime,q^\prime).
\ee
The time reversibility condition means the following is satisfied.
\be 
T_{MD}(-\Delta t):(p^\prime,q^\prime)\rightarrow (p,q).
\ee
The simplecticity is the condition that the phase space volume is conserved, 
$dpdq=dp^\prime dq^\prime$. 
The symplectic integrators in Sec.2 satisfy the above two conditions.

This elementary MD step is performed repeatedly 
to integrate the equations up to a certain ``length'' of the MD trajectory.
Here the trajectory length is set to unity.
In general the integrator can not solve the Hamilton's equations of motion exactly
and thus the energy is not conserved.  
Let $\Delta H$ be the change in energy caused by the integrator at the end of the trajectory.
This error is corrected by the Metropolis test, which accepts the candidate configuration $\{U'\}$ at the end of the trajectory 
with probability $p=\min(\exp(-\Delta H),1)$.
If the candidate configuration is rejected, the old configuration $\{U\}$ is included
again in the Markov chain.

Thus, a trade-off must be achieved between two conflicting goals:
good energy conservation, because it ensures high acceptance probability in 
the Metropolis step, and low computer cost.
This is why the choice of integrator plays a crucial role.

The fermionic part of the force $\displaystyle \partial H/\partial U$
is given by 
\be
\frac{\partial H_{fermionic}}{\partial U} =-\phi^\dagger (MM^\dagger)^{-1}\frac{\partial (MM^\dagger)}{\partial U}
(MM^\dagger)^{-1}\phi.
\ee
Since the matrix $M$ is singular at the quark mass $m=0$, as $m$ decreases the fermionic force diverges.
Thus HMC simulations in lattice QCD become unstable at small quark masses, 
unless the stepsize is reduced in proportion to $m$.  

\subsection{Error of Hamiltonian and Acceptance}
Here we summarize the expected behavior for the error of the Hamiltonian and the acceptance 
of the HMC algorithm.
The $n$-th order integrator causes ${\cal O}(\DT^{n+1})$ integration errors for $q$ and $p$.
However the error in the Hamiltonian at the end of a unit-time trajectory\footnote{
$\Delta H$ does not increase linearly with trajectory length.
It increases linearly up to a certain characteristic length $l_c$ 
provided that $\DT$ is not too large, then saturates.
Thus the accumulated error in the Hamiltonian is expected to be 
$\Delta H  \sim {\cal O}(\DT^{n+1}) \times \frac{l_c}{\DT}
\sim {\cal O}(\DT^{n})$\cite{Gupta:1990ka}.}
 is ${\cal O}(\DT^n)$.
Thus $\Delta H \sim \DT^n$.
Furthermore from Creutz's equality $\bra \exp(\Delta H)\ket=1$ \cite{Creutz:1988wv} 
we expect 
\be 
\bra\Delta H\ket\sim \bra\Delta H^2\ket \propto V\DT^{2n},
\ee
where $V$ is the volume of the system.
Thus the root mean square of  the error of the Hamiltonian 
at small $\DT$ is expected to be
\be
 \bra\Delta H^2\ket^{1/2} \approx C_n V^{1/2} \DT^n ,
\label{eq:DH}
\ee
where $C_n$ is a Hamiltonian-dependent coefficient.
Using $\bra\Delta H^2\ket^{1/2}$, the acceptance of the HMC algorithm 
for large volumes is given by \cite{Gupta:1990ka}
\be 
\bra P_{acc}\ket={\rm erfc}(\bra\frac18 \Delta H^2\ket^{1/2}).
\label{eq:acc1}
\ee
For small $\bra\Delta H^2\ket^{1/2}$, one may use 
the approximate formula:
\be
\bra P_{acc}\ket=\exp(-\frac1{\sqrt{2\pi}}\bra \Delta H^2\ket^{1/2}),
\label{eq:acc2}
\ee
which is applicable for $P_{acc} \ge 20\%$\cite{Takaishi1}.
The performance of integrators can be measured by 
the [inverse of the] work per accepted trajectory, i.e.
by the product of the acceptance and step size: $P_{acc}\times \DT$.
The best performance of integrators is obtained  at  
the step size which maximize $P_{acc}\times \DT$.
Using eqs.(\ref{eq:DH}) and (\ref{eq:acc2}) 
we obtain the optimal acceptance which maximizes  $P_{acc}\times \DT$ as \cite{Takaishi1}
\be
P_{opt} = \exp\left(-\frac1n\right),
\label{eq:opt}
\ee
which does not depend on the details of the Hamiltonian 
but only on the order of the integrator.
This result indicates that the optimal acceptance for any 2nd order integrator 
is about 61\% which is consistent with the numerical results of $60 \sim 70\%$\cite{Takaishi1}.
Eq.(\ref{eq:opt}) also indicates that the optimal acceptance increases with the order of 
the integrator: 78\% for 4th order and 85\% for 6th order.

\subsection{Performance of 2nd order MN (2MN) integrator}
Here we compare the efficiency of the 2MN integrator with that of the 2LF integrator.
For 2nd order  integrators, from eq.(\ref{eq:DH}) $\bra\Delta H^2\ket^{1/2}$ at small $\DT$ is expected to be 
$C_2 V^{1/2} \DT^2$. 
We measure the coefficient $C_2$ for both integrators at small enough $\DT$,
and by comparing the coefficients we obtain the performance of the 2MN integrator
relative to the 2LF integrator.

\begin{figure}
\vspace{5mm}
\begin{center}
\epsfig{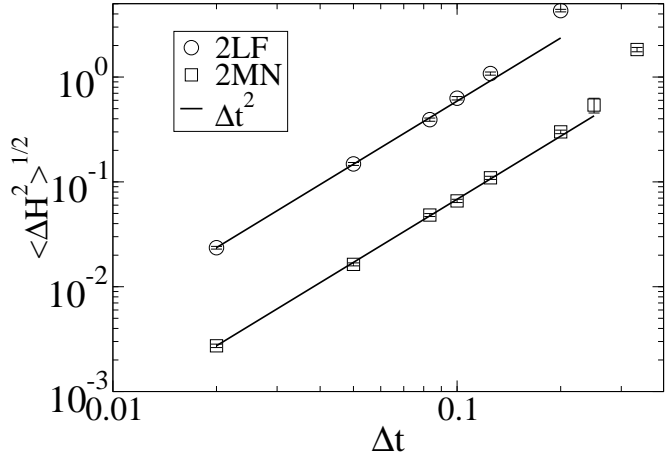}
\caption{\label{fig:dhvsdt}  $\bra\Delta H^2\ket^{1/2}$  as a function of $\DT$. 
Simulations are performed at $\beta=5.00$ and $\kappa = 0.160$ on  $4^4$ lattices.
The line proportional to $\DT^2$ is drawn to guide the eye.
}
\end{center}
\end{figure}

Figure \ref{fig:dhvsdt} shows $\bra\Delta H^2\ket^{1/2}$  as a function of step size $\DT$
at $\beta=5.00$ and $\kappa=0.160$ on $4^4$ lattices. We see that $\bra\Delta H^2\ket^{1/2}$ is proportional 
to $\DT^2$ as expected and the error of the 2MN integrator is about 10 times smaller than that of 
the 2LF integrator at any $\DT$ until instabilities show up.

\begin{figure}
\vspace{5mm}
\begin{center}
\epsfig{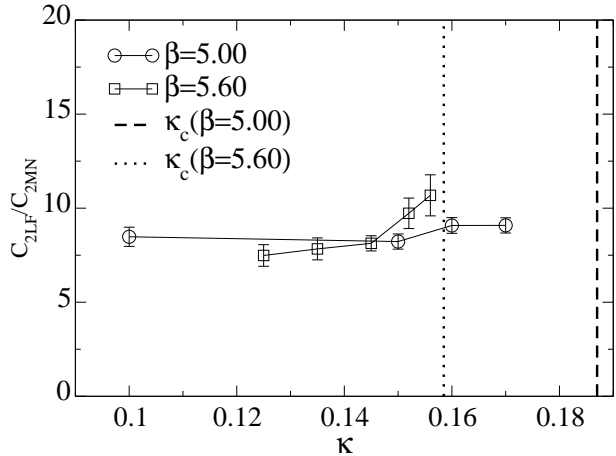}
\caption{\label{fig:ratio} $C_{2LF}/C_{2MN}$ as a function of $\kappa$. Simulations at $\beta=5.00(5.60)$ are performed on  $4^4(8^4)$ lattices.
}
\end{center}
\end{figure}

Figure \ref{fig:ratio} shows the ratio $C_{2LF}/C_{2MN}$ as a function of $\kappa$.
The coefficients $C_{2LF}$ and $C_{2MN}$ are extracted by using eq.(\ref{eq:DH}) for the 2nd order
with simulations at a small value of $\DT$.
As seen in the figure, $C_{2LF}/C_{2MN}$ is about 10, which means that the error of the 2MN integrator 
is about 10 times smaller than that of the 2LF integrator. 
This is consistent with the theoretical expectation of Omelyan {\it et al.}.
If we take $C_{2LF}/C_{2MN}\approx 10$, 
this means that the step size of the 2MN integrator  can be increased 
by a factor 3 ($\approx\sqrt{10}$) over that of the 2LF integrator,
as long as the error still behaves as $\DT^2$. 
Since the 2MN integrator has two force calculations per elementary step, 
the efficiency should be measured by $\displaystyle \sqrt{C_{2LF}/C_{2MN}}/2$, 
which is about 1.5. 
Thus it is concluded that the 2MN integrator is about 50\% faster than the 2LF integrator.

Of course, this assessment rests on the assumption that the step size can indeed be
increased without running into instabilities, so that the limiting factor in the 
step size comes from the error accumulation. Note that the 2MN integrator appears no
worse, or perhaps slightly better, than the 2LF in terms of instabilities: departure from
the quadratic behaviour $\bra\Delta H^2\ket^{1/2} \propto \DT^2$ starts at similar values of 
$\bra\Delta H^2\ket^{1/2}$ in Fig.~\ref{fig:dhvsdt}, and appears more gradual.

\subsection{Comparison of 2nd and 4th order MN integrators }
\begin{figure}
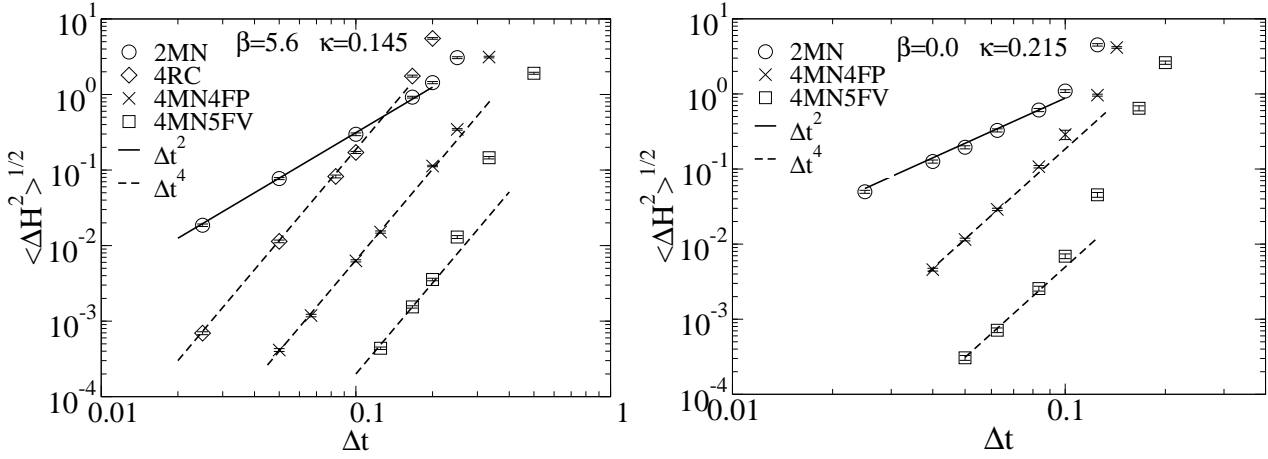

\vspace{5mm}
\begin{center}
\epsfig{file=compdHb560k145-2.eps,height=6cm}
\epsfig{file=compdHb00k215-2.eps,height=6cm}
\caption{\label{fig:compdH}$\bra\Delta H^2\ket^{1/2}$ as a function of $\DT$. Simulations are performed on $8^4$ lattices.
4RC stands for the 4th order integrator obtained by eq.(\ref{eq:rec}).
The lines proportional to $\DT^n$ are drawn to guide the eye.
}
\end{center}
\end{figure}

The efficiency of higher order integrators should be measured against lower order ones.
From the above analysis we know that the 2MN integrator is more efficient than 
the 2LF integrator.
Therefore we compare the 4MN integrator with the 2MN integrator.
Assuming eq.(\ref{eq:DH}) the comparison could be done by following
the analysis of \cite{Takaishi1}.
However we found a problem with the 4MN integrator.
Namely the error of the Hamiltonian $\bra\Delta H^2\ket^{1/2}$ is not simply described by eq.(\ref{eq:DH}) 
but is dominated by higher order terms in $\DT$
already at small  $\bra\Delta H^2\ket^{1/2}$.
Figure \ref{fig:compdH} shows $\bra\Delta H^2\ket^{1/2}$ on $8^4$ lattices as a function of $\DT$.
As seen in the figure (left), at a fixed step size the error of the 4MN5FV integrator 
is about 1000 times smaller
than the previously known 4th order integrator (4RC), 
which is consistent with the theoretical expectation\cite{Omelyan}.
The expected behavior of $\bra\Delta H^2\ket^{1/2} \approx C_4 V^{1/2} \DT^4$, however,
is seen only at small $\bra\Delta H^2\ket^{1/2}$.
We are only interested in the region of $0.1 \le \bra\Delta H^2\ket^{1/2} \le 1$ which corresponds to 
an acceptance of $60\% \sim 95\%$\footnote{Figure 1 of \cite{Takaishi1}}.
In this region, $\bra\Delta H^2\ket^{1/2}$ of the  4MN5FV integrator is dominated by higher order terms in $\DT$,
which results in that $\bra\Delta H^2\ket^{1/2}$ grows rapidly with $\DT$.
This observation makes the 4MN integrator unattractive on practical
lattice sizes.

Although the 4MN4FP integrator seems to be more stable than the 4MN5FV integrator,  
it also shows the deviation from the $\DT^4$ line at small $\bra\Delta H^2\ket^{1/2}$ (See Fig.3).
Thus compared to the 2MN integrator, 
the 4MN integrators tested here seem unattractive.
As the quark mass $m$ decreases, it is often seen that the integrator becomes unstable\cite{Kennedy}, because the force increases as $1/m$.
In lattice QCD calculations the parameter region of small quark masses
is the physically interesting one.  
In this region the 4MN integrators may easily show instability, 
which limits their applicability.

At finite temperature, however, the coefficients $C_n$ behave differently from those at zero temperature.
Typically we expect $C_n^{T\neq 0} \leq C_n^{T=0}$.  
Therefore at finite temperature we may be lead to a different conclusion and this must be studied numerically.  
A numerical test showed that at finite temperature the 4RC integrator performs 
better than the 2LF integrator on lattices larger than a minimum size\cite{Takaishi2}.
We have made the same test for the 2MN and 4MN5FV integrators 
on an $18^3\times 4$ lattice at $\beta =5.75$ and $\kappa=0.1525$. 
For the 2MN integrator the acceptance is measured to be about 0.6 at $\Delta t =0.1$ and 
for the 4MN5FV integrator the acceptance is about 0.8 at $\Delta t =0.37$.
These values of the acceptance are close enough to the optimal acceptance given by eq.(\ref{eq:opt}).
The gain of the 4MN5FV integrator over the 2MN one is calculated by
\be
G= \frac{(P_{acc}\times \Delta t)_{4th}}{\kappa_{42}(P_{acc}\times \Delta t)_{2nd}},
\ee
where $\kappa_{42}$ is the relative cost factor and 
$\kappa_{42}=2.5$ for the 2MN and 4MN5FV integrators.
Substituting the measured values into $G$, $G$ is calculated to be $\approx 2$,
which shows that the 4MN5FV integrator is more effective.   
Thus at finite temperature there is room to use a 4MN integrator depending on the simulation parameters. 

\section{Tuning the 2MN integrator}
The strategy to minimize eqs.(\ref{eq:Err3}) and (\ref{eq:Err5}) is based on the assumption 
that the errors coming from the two commutators $[T,[V,T]]$ and $[V,[V,T]]$ are 
equally dominant.
In general this simplifying assumption does not hold.
Here we try to minimize a more general form of the error function. 
Let us assume the following form of $\bra\Delta H^2\ket^{1/2}$,
\bea
 \bra\Delta H^2\ket^{1/2} & \approx &  \sqrt{\alpha(\lambda)^2 f(\DT)^2 + \beta(\lambda)^2 g(\DT)^2}, \\
& = &  \sqrt{\alpha(\lambda)^2 f^2 + \beta(\lambda)^2 g^2}\DT^2,
\label{eq:norm}
\eea
where $\alpha(\lambda)$ and $\beta(\lambda)$ are given by eqs.(\ref{eq:alpha_lambda}) and (\ref{eq:beta_lambda}),
and $f^2$ and $g^2$ are unknown parameters, to be determined from numerical simulations.
In general, by performing simulations at two values of $\lambda$ one can determine $f^2$ and $g^2$ numerically.
The determination can be made easier by noticing that 
$\alpha(\lambda_1)$ and $\beta(\lambda_2)$  are zero at $\lambda_1=(1-1/\sqrt{3})/2$ and 
$\lambda_2=1/6$ respectively. By simulating at $\lambda_1$ and $\lambda_2$ 
we immediately obtain $f^2$ and $g^2$.

\begin{figure}
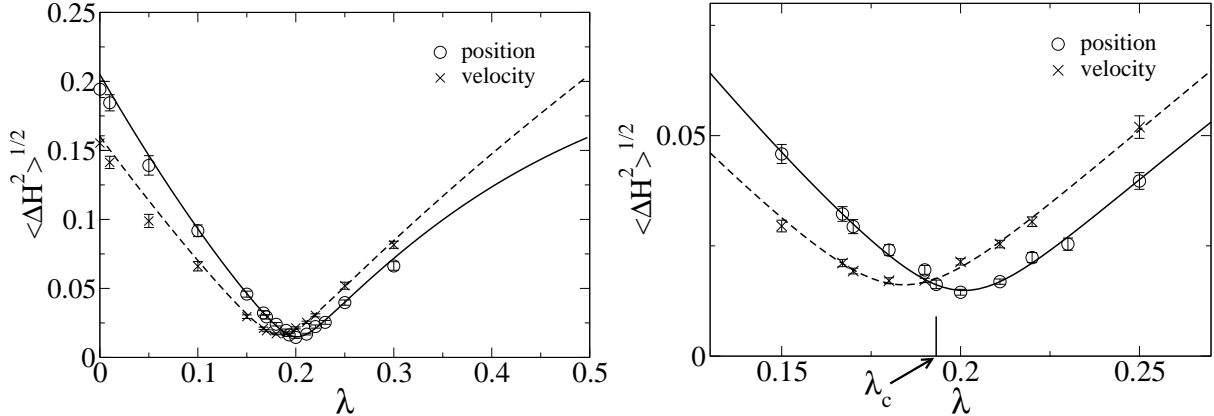

\vspace{5mm}
\begin{center}
\epsfig{file=normb500k160.eps,height=5.5cm}
\epsfig{file=normb500k160-2.eps,height=5.5cm}
\caption{\label{fig:norm} $\bra\Delta H^2\ket^{1/2}$  as a function of 
$\lambda$. The right figure is a zoom of the left. 
Simulations are performed at $\beta=5.00$ and $\kappa=0.160$ on  $4^4$ lattices with $\DT=0.05$.
The lines are determined from simulations of the position version integrator at $\lambda_1$ and $\lambda_2$.
The position version has a small advantage over the velocity version, since
it gives a slightly reduced minimum RMS error (right).
}
\end{center}
\end{figure}

Figure \ref{fig:norm} shows $\bra\Delta H^2\ket^{1/2}$ at $\beta=5.00$ and $\kappa=0.160$
as a function of $\lambda$. We see that the optimal $\lambda$ at the minimum of $\bra\Delta H^2\ket^{1/2}$
is slightly different from $\lambda_c$, the value of eq.(\ref{eq:optlambda}).
Moreover the optimal $\lambda$ is different between the velocity and position version integrators.
Here eq.(\ref{eq:2MN}) is the position version integrator. 
The velocity version is obtained by interchanging $T$ and $V$ in  eq.(\ref{eq:2MN}). 
The lines in the figure are given by eq.(\ref{eq:norm}), with $f^2$ and $g^2$ determined by simulations 
of the position version of 2MN integrator at $\lambda_1$ and $\lambda_2$.
To draw the dashed line for the velocity version, we simply interchange the values $f^2$ and $g^2$.
Both lines describe the numerical results very well, down to $\lambda=0$.
Note that the velocity version of 2MN integrator becomes the position version of 2LF integrator at $\lambda=0$, and vice versa.
The position version of the 2MN integrator gives a slightly smaller minimum. 
At $\lambda=0$ the velocity version of 2MN integrator has 
a smaller error than the position version, which means that 
the position version of 2LF integrator has a smaller error than the velocity version.  
This was already observed in \cite{Kilcup}.
Since the position version also leads to one less force evaluation by the end of
a trajectory, we definitely recommend using the position version (for the 2MN and the
2LF integrators both): it requires less work and gives a higher acceptance.

The quality of our fit justifies a posteriori the ansatz made for the
magnitude of the error eq.(\ref{eq:norm}). Indeed, to leading order $\Delta t^4$, the
error $\bra\Delta H^2\ket$ should be of the form
\be
\bra\Delta H^2\ket = (\alpha(\lambda)^2\bra F^2\ket +\alpha(\lambda)\beta(\lambda)\bra FG + GF\ket
+\beta(\lambda)^2\bra G^2\ket)\Delta t^4,
\ee
where $F=[T,[V,T]]$ and $G=[V,[V,T]]$ from eq.(\ref{eq:alpha}). We find that the crossterm
$\bra FG\ket$ is in our case "one order of magnitude smaller" 
than $\bra F^2\ket$ and $\bra G^2\ket$, indicating that the two operators are almost
uncorrelated in our system. While this finding may not be true in general,
it provides support for the minimum norm strategy of Omelyan et al.
at least in the context of lattice QCD.

\begin{figure}
\vspace{5mm}
\begin{center}
\epsfig{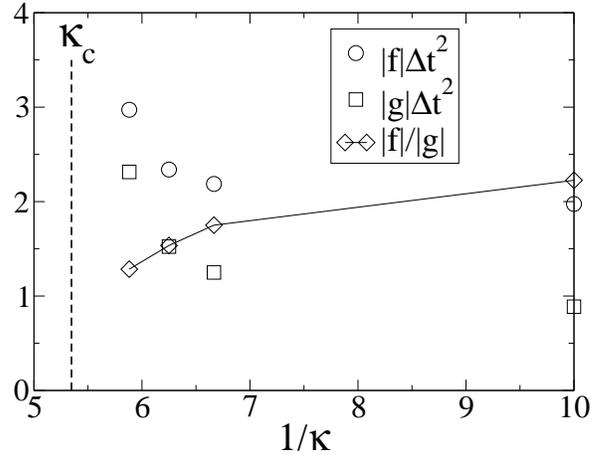}
\caption{\label{fig:a1b1} $|f|\DT^2$, $|g|\DT^2$ and $|f|/|g|$ as a function of $1/\kappa$.
Simulations are performed at $\beta=5.00$ on $4^4$ lattices with $\DT=0.05$.
The dashed line indicates $\kappa_c =0.187$\cite{Ukawa}.
}
\end{center}
\end{figure}

Figure \ref{fig:a1b1} shows $|f|\DT^2$, $|g|\DT^2$ and $|f|/|g|$ as a function of $1/\kappa$ 
at $\beta=5.00$ on $4^4$ lattices ($\DT=0.05)$. 
As one approaches $\kappa_c$, both $|f|$ and $|g|$ increase.
On the other hand the ratio  $|f|/|g|$ decreases. 
This is as expected since $g$ comes from the error term $[V,[V,T]]$ involving two factors 
of the fermion potential, versus one for $f$ and $[T,[V,T]]$.
Although $|f|/|g|$ seems to approach one at $\kappa_c$, there is a possibility that 
it further goes down to zero and its limit must be carefully investigated.
Note that when $|f|/|g|=1$, the optimal $\lambda$ becomes $\lambda_c$.

\begin{figure}
\vspace{5mm}
\begin{center}
\epsfig{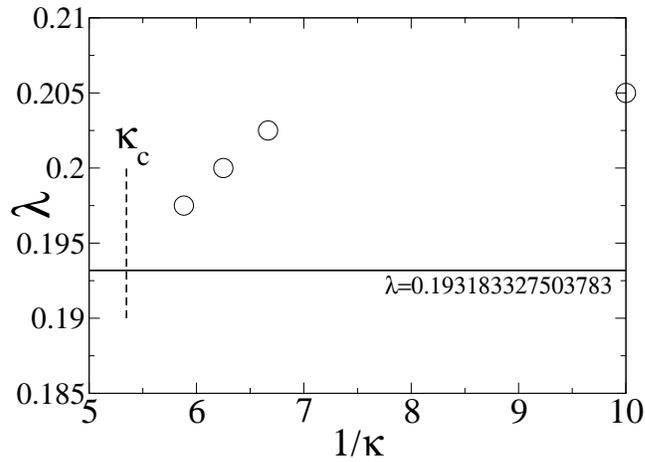}
\caption{\label{fig:optlambda} Optimal $\lambda$ as a function of $1/\kappa$. 
Simulations are performed at $\beta=5.00$ on $4^4$ lattices with $\DT=0.05$.
The dashed line indicates $\kappa_c =0.187$.
}
\end{center}
\end{figure}

Figure  \ref{fig:optlambda} shows the optimal $\lambda$ as a function of $1/\kappa$.
We see that the optimal $\lambda$ is different from $\lambda_c$ and slightly larger.

\section{Conclusions}
We have tested the new 2nd and 4th order integrators obtained by minimizing the norm of the error coefficients.
We find that the 2MN integrator performs better than the conventional 2LF integrator, by about 50\%.
Therefore we recommend to use the 2MN integrator in HMC simulations. 
Although in our tests we used the standard Wilson fermion action, 
the 2MN integrator can be used for any actions, e.g, KS fermions, improved actions, polynomial actions 
for odd flavors\cite{Takaishi:2001um,deForcrand:1996ck}.  
Moreover we may combine the 2MN integrator with other acceleration techniques such as 
multiple time step integration\cite{Sexton:1992nu}, 
multiple pseudo-fermions\cite{Hasenbusch:2001ne} 
and preconditioned actions\cite{deForcrand:1996ck}.

The same 2MN integrator can also be used in the Trotter-Suzuki
decomposition of the partition function: $\exp(-\beta H) = (\exp(-\DT~H))^N$,
where $N = \beta/\DT$, in Quantum Monte Carlo simulations, when a formulation
in continuous imaginary time \cite{Wiese} is not practical.

Although at first sight, one can equivalently start by integrating over positions
or velocities, we observe that integrating over positions first gives a slightly
higher acceptance \cite{Kilcup}, with one less force evaluation at the end of a
trajectory.

\section*{Acknowledgements}
The authors thank the Institute of Statistical Mathematics for the use of NEC SX-6, 
RCNP at Osaka University for the use of NEC SX-5 and the Yukawa Institute 
for the use of NEC SX-5, and Tony Kennedy for his comments on the manuscript.
T.~T. would like to thank Prof.~Sigrist for hospitality during his stay at ETH Z\"urich.
P. de F. thanks the Kavli Institute for Theoretical Physics for hospitality
during the completion of this paper.

\end{document}